\documentclass[11pt]{article}

\usepackage[final]{acl}

\usepackage{times}
\usepackage{latexsym}

\usepackage[T1]{fontenc}

\usepackage[utf8]{inputenc}

\usepackage{microtype}

\usepackage{inconsolata}

\usepackage{graphicx}

\usepackage{hyperref}
\usepackage{url}
\usepackage{graphicx}
\usepackage{wrapfig}
\usepackage{booktabs}
\usepackage{multirow}
\usepackage{array}
\usepackage{siunitx}
\usepackage{xcolor}
\usepackage{colortbl,booktabs}
\usepackage{enumitem}
\usepackage{pifont}
\usepackage[most]{tcolorbox}
\usepackage{hyperref}
\usepackage{arydshln}

\usepackage{amsmath}      
\usepackage{amssymb}
\usepackage{algorithm}
\usepackage{algpseudocode}
\usepackage{cancel}
\usepackage{circledsteps}
\usepackage{float}

\newtheorem{definition}{Definition}
\title{Generate What You Can Trust: Content Credibility in \\Generative Recommenders}

\author{
  Zhuo Cai\textsuperscript{1} \quad
  Guanghao Wu\textsuperscript{1} \quad
  Shoujin Wang\textsuperscript{1}\thanks{Corresponding author: Shoujin.Wang@uts.edu.au} \\
  \textbf{Peilin Zhou\textsuperscript{2}} \quad
  \textbf{Victor W. Chu\textsuperscript{1}} \\
  \textsuperscript{1} Data Science Institute, University of Technology Sydney, Australia \\
  \textsuperscript{2} New York University Abu Dhabi, United Arab Emirates \\
  \texttt{\{Zhuo.Cai, Guanghao.Wu\}@student.uts.edu.au}, \texttt{Shoujin.Wang@uts.edu.au}, \\ \texttt{zhoupalin@gmail.com},  \texttt{WingYan.Chu@uts.edu.au}
}

\begin{document}
\maketitle
\begin{abstract}
Generative recommendation (GR) represents items with semantic IDs (i.e., discrete token sequences) and generates target item tokens as recommendations. Despite its promising results, existing methods predominantly optimize for accuracy while neglecting the credibility of the recommendations they generate. This oversight inevitably exposes users to uncredible content (e.g., fake news) with serious societal consequences, including user distrust, reputation harm to platforms, and broader social instability. To address this critical yet underexplored challenge, we propose \texttt{CreGR}, the first credible GR model that jointly tackles content credibility across the two core stages of GR: \textit{tokenization} and \textit{generation}. In the tokenization stage, we design a new \textbf{credibility-aware tokenizer} that explicitly encourages the model to learn discriminative tokens respectively for credible and uncredible items, thereby disentangling credibility signals at the token level. Building on this, in the generation stage, we propose a novel \textbf{accuracy-preserving and credibility-oriented generator} grounded in discrete diffusion. Specifically, we introduce an asymmetric masking probability reduction strategy that selectively diminishes the contribution of tokens associated with uncredible content to the generation process, while leaving tokens encoding user preference signals unaffected so as to preserve recommendation accuracy. Experiments on three real-world datasets demonstrate the effectiveness of \texttt{CreGR}. The implementation code and preprocessed datasets of our method are provided in \url{https://github.com/iamZhuoCai/CreGR}.
\end{abstract}

\section{Introduction}
\begin{figure}[t]
    \centering
    \includegraphics[width=1\linewidth]{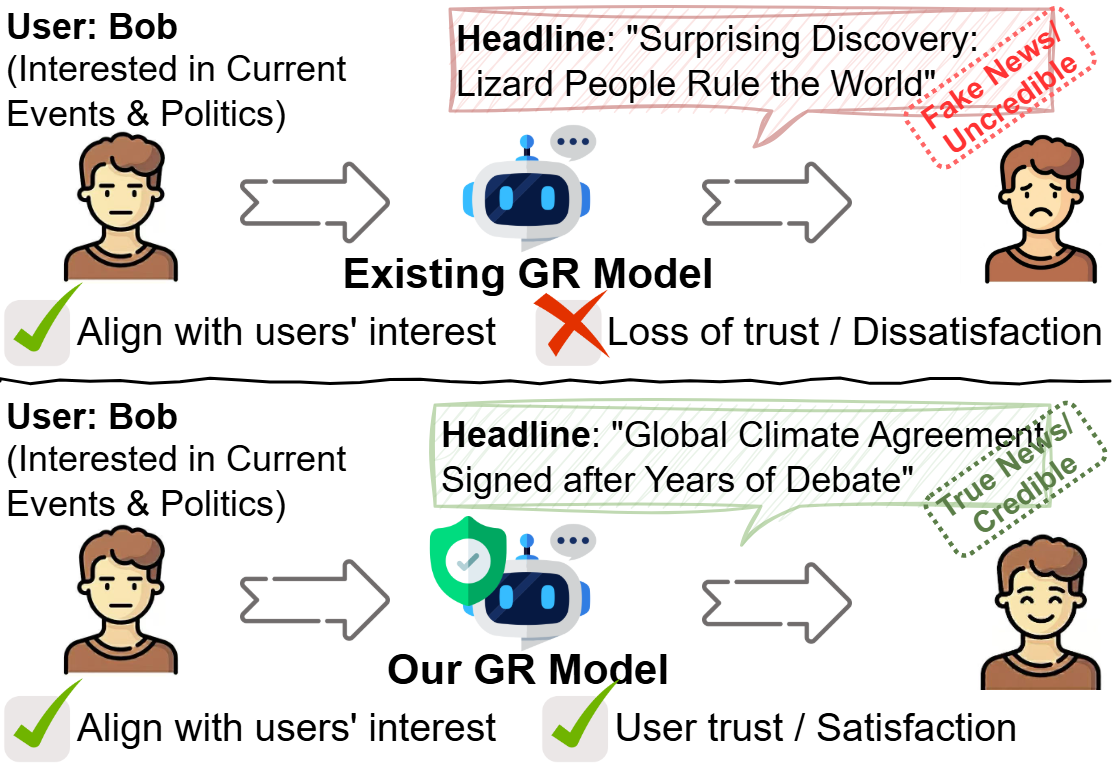}
    \caption{Comparison between existing GR models and our proposed credible GR model. Existing GR models generally neglect the content credibility of recommendations. Our proposed model is capable to improve recommendation credibility while preserving high recommendation accuracy.}
    \label{fig:toy_example}
\end{figure}

Generative recommendation~(GR) has emerged as a transformative paradigm for recommender systems in today's generative AI era, reformulating the recommendation problem from conventional embedding-based matching to token generation~\cite{rajput2023recommender, 
wang2024learnable, hou2025generating}. Compared to traditional methods, GR offers superior generalization that subsumes retrieval and ranking into a single generation
process. Despite its remarkable success and rapid adoption by major platforms serving hundreds of millions of users \cite{deng2025onerec, singh2024better, he2026plum, penha2025semantic}, existing GR methods are predominantly optimized for recommendation accuracy while neglecting a critical yet underexplored societal dimension: the \textbf{content credibility} of the generated recommendations. As illustrated in Figure~\ref{fig:toy_example}, existing GR methods may accurately capture user preference yet recommend uncredible content (e.g., fake news) to users. This oversight leads to serious consequences, including eroded user trust, reputational risks to platforms, and even social instability \cite{cai2026steering, wang2022trustworthy, wang2022data, zhang2023rumor}.

There are some recent efforts on developing credible content recommendation methods \cite{wang2022veracity,wang2024hierarchical, cai2026steering, cai2026misinformation, sallami2023trust}. For example, some works \cite{wang2022veracity,wang2024hierarchical} disentangle news content into veracity embeddings and event embeddings, and utilize only event embeddings to compute user-item matching scores, thereby mitigating the influence of uncredible content. Cai et al. \cite{cai2026steering} employ a diffusion model to improve recommendation credibility. However, these methods all operate within the \emph{embedding matching} paradigm, where recommendations are made by computing similarity scores between user and item embeddings over a large candidate pool. As such, they are inherently incompatible with the emerging GR paradigm, which forgoes embedding matching entirely and instead directly generates the semantic IDs~(SIDs) of target items. In this paper, we focus on how to improve the credibility of GR. However, building a credible GR model is non-trivial, primarily due to the following two challenges (CHs):

\textbf{CH1: How to learn discriminative semantic IDs for credible and uncredible content items in the tokenization stage?}
Existing methods tokenize items by encoding their metadata (e.g., titles and descriptions) through pre-trained language models and subsequently quantizing the resulting embeddings into discrete tokens (i.e., SIDs) via vector quantization (e.g., RQ-VAE \cite{lee2022autoregressive}). However, these models are pre-trained with objectives such as masked language modeling or next-token prediction, which capture topical and syntactic patterns but are inherently agnostic to content credibility~\cite{chuang2024dola, lin2022truthfulqa}. As a result, credible and uncredible content that share similar topics are mapped to nearby regions in the embedding space. This credibility entanglement causes topically similar but factually divergent items to be assigned similar/nearby tokens, making them difficult to discriminate during the subsequent generation process and potentially leading to the recommendation of uncredible content.

\textbf{CH2: How to balance recommendation accuracy and credibility in the generation stage?}
A straightforward solution to mitigate uncredible content is to directly remove uncredible items from training data and retrain the generation model. However, this naive removal strategy inevitably degrades recommendation accuracy, as the valuable information (e.g., topics) in the uncredible items that reflect users' preference is also removed. For instance, a user who frequently engages with uncredible health news articles still reveals a strong interest in wellness topics, which should be leveraged to capture the user's preference and support accurate generation. The empirical evidence is provided in Appendix \ref{simply_remove}. Therefore, there is a fundamental trade-off between recommendation accuracy and credibility in the generation stage.

To address these challenges, in this paper, we propose a novel \underline{cre}dible \underline{g}enerative \underline{r}ecommendation model, \texttt{CreGR}. Specifically, to overcome \textbf{CH1}, we design a novel credibility-aware tokenizer that introduces credibility regularization to enforce intra-class compactness and inter-class separation between credible and uncredible items. This compels the codebook to disentangle credibility signals at the discrete token level and thereby eliminating entanglement between factually divergent items. To address \textbf{CH2}, we propose an accuracy-preserving and credibility-oriented generator grounded in discrete diffusion. Exploiting the fact that discrete diffusion training is conditioned on masked tokens, we design an asymmetric masking probability reduction strategy that selectively reduces the masking probabilities of tokens associated with uncredible content identified in the tokenization stage, while leaving all other 
tokens encoding valuable user preference signals at their original masking probabilities. We instantiate this strategy through two variants: a \textit{hard} weighting strategy that performs absolute erasure of uncredible token contributions, 
and a \textit{soft} weighting strategy that applies distance-aware down-weighting based on proximity to cluster centroids. These designs gracefully suppress uncredible generation while preserving the topical preference signals, achieving a balance between recommendation accuracy and credibility.

Overall, our contributions are summarized as:
\begin{itemize}[leftmargin=*, itemsep=2pt, topsep=2pt, parsep=0pt, partopsep=2pt]
    \item We propose a novel credible generative content recommendation model, \texttt{CreGR}, which is the first GR model that considers the credibility of the generated recommendations.
    \item We design a credibility-aware tokenizer and a novel accuracy-preserving and credibility-oriented generator, which work collaboratively to improve the recommendation credibility while preserving high recommendation accuracy.
    \item Extensive experiments on three real-world datasets demonstrate the effectiveness of our proposed model in generating credible and accurate content recommendations.
\end{itemize}

\section{Problem Formulation}
\subsection{Credible Content Recommendation}
Following prior work \citep{wang2022veracity, wang2024hierarchical, cai2026steering}, we formulate credible content recommendation as the task of reducing user exposure to uncredible items, such as fake news and misinformation, which degrade user experience and pose adverse societal impacts. Credible content recommendation follows the task of sequential recommendation \cite{wang2021survey, wang2019sequential, wang2019modeling, wang2018attention, cai2025unleashing}. A recommendation model is deemed more credible when its generated recommendation lists contain fewer such items.

Besides, \textit{this task requires developing a recommendation model that inherently ensures the credibility, rather than relying on post-hoc filtering} (i.e., removing identified uncredible items from final recommendation lists). Post-hoc filtering treats the symptom rather than the cause: the underlying model has already learned latent patterns favoring uncredible content during training, and such filtering remains vulnerable to adversarial manipulation where uncredible content is crafted to evade detection \cite{zhou2020survey}.

\begin{definition}
\textbf{Content credibility.} Content credibility is defined as the trustworthiness of the information conveyed by an item. An item is regarded as uncredible if it contains false, misleading, or inaccurate information (e.g., fake news, health misinformation), and credible otherwise.
\end{definition}

\begin{figure*}
    \centering
    \includegraphics[width=1\linewidth]{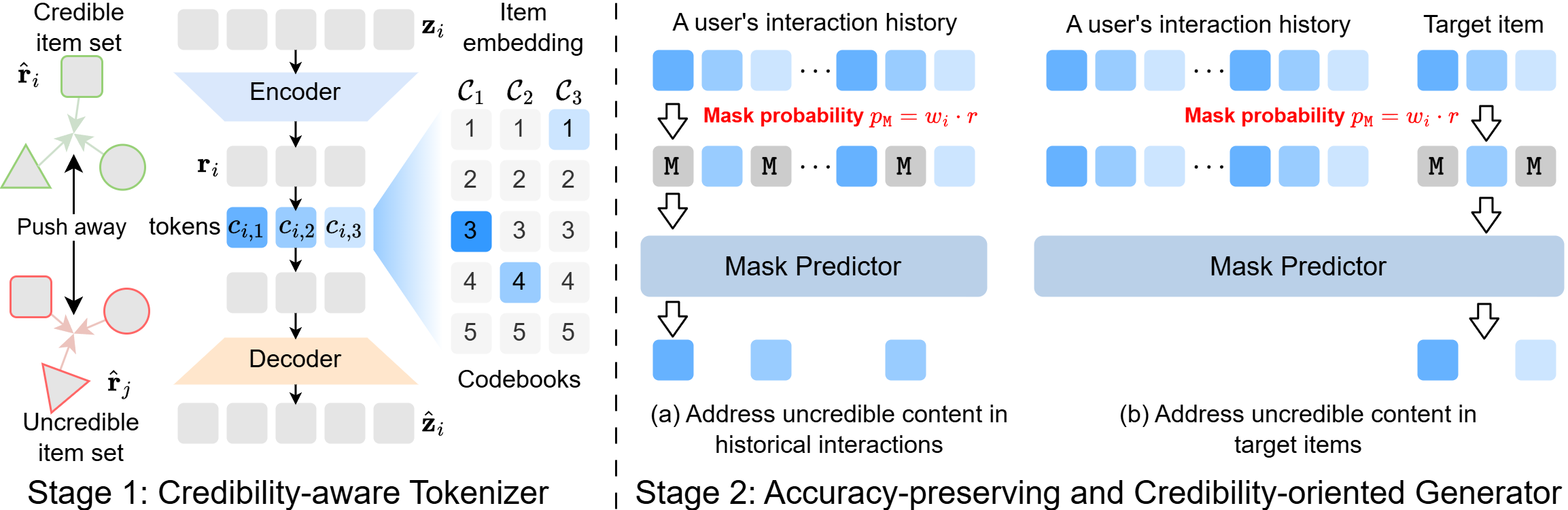}
    \caption{Overall Framework of our proposed model. It consists of two stages: (1) a credibility-aware tokenizer, which tokenizes items into discrete tokens while also disentangling credible and uncredible items in token level;  and (2) an accuracy-preserving and credibility-oriented generator, which mitigates the generation of uncredible content items while preserving high recommendaiton accuracy.}
    \label{fig:framework}
\end{figure*}

\subsection{Generative Recommendation}
Given a user's interaction history $\mathcal{H} = \{v_1, v_2, \dots, v_{n-1}\}$, GR formulates the recommendation task as a token generation problem to generate the tokens of the target item $v_{n}$ that users may be interested in. Each item $v_i$ is represented as a sequence of discrete tokens (i.e., SIDs) $[c_{i,1}, c_{i,2}, \dots, c_{i,L}]$ using vector quantization techniques (e.g., RQ-VAE), where $L$ denotes the number of quantization levels. Afterwards, the user's interaction sequence can be represented as:
\begin{equation}
      \mathcal{S}_{\mathcal{H}}=[c_{1,1}, \cdots, c_{1,L}; \cdots; c_{n-1,1}, \cdots, c_{n-1,L}].
\end{equation}

GR aims to learn a probabilistic model $\mathcal{P}_\theta$ that generates the tokens of the target item $v_{n}=[c_{n,1}, \cdots,c_{n,L}]$ conditioned on $\mathcal{S}_{\mathcal{H}}$: $\mathcal{P}_\theta(v_{n}|\mathcal{S}_\mathcal{H})$.

Existing GR models can be categorized into autoregressive (AR) models that generate the tokens of target items from left-to-right, and non-autoregressive (NAR) models that generate tokens in parallel via discrete diffusion. Our proposed method is based on the second category.


\noindent\textbf{NAR GR.} This paradigm generates the tokens of the target item in parallel via discrete diffusion modeling \cite{shi2025llada}. It operates through a forward masking process and a reverse denoising process. In the forward process, each token of the target item is independently masked with mask token $\mathtt{[M]}$ with probability $r$, or remains unchanged with probability $1-r$. In the reverse process, the model $\mathcal{P}_{\theta}$ predicts
all masked tokens simultaneously. Specifically, the reverse process proceeds over $T$ steps. At each step $t$, the model predicts the original tokens for all currently masked positions. Formally, $\mathcal{P}_\theta(v_{n}|\mathcal{S}_{\mathcal{H}})$ can be factorized as:
\begin{equation}
        \prod_{t=1}^{T} \prod_{l=1}^{L}
    \begin{cases}
        \mathcal{P}_\theta(c_{n,l}|v_{n}^t, \mathcal{S}_{\mathcal{H}}), & \text{if } c_{n,l}^t = \mathtt{[M]}, \\
        1, & \text{otherwise},
    \end{cases}
\end{equation}
where $v_{n}^t$ is the generated item token sequence at step $t$, and $c_{n,l}^t$ denotes the generated token at level $l$ and step $t$. At step $t$, the model selects the top-$\frac{L}{T}$ predictions to unmask, and the remaining positions are re-masked for subsequent steps.

\section{Methodology}
As shown in Figure~\ref{fig:framework}, \texttt{CreGR} comprises two key components. \textbf{(1) Credibility-aware tokenizer} discretizes each item into token sequences while enforcing discriminative representations between credible and uncredible items at the token level. \textbf{(2) Accuracy-preserving and credibility-oriented generator} reduces the masking probabilities of tokens encoding uncredible information identified by the tokenizer, while leaving all other tokens unaffected, thereby suppressing the influence of uncredible content in the generation process without compromising recommendation accuracy.
\subsection{Credibility-aware Tokenizer}
We first employ LLaMA2-7B~\cite{touvron2023llama} to encode the title and description of each item $v_i$ into a title embedding $\mathbf{e}_{i}^{tit}$ and a description embedding $\mathbf{e}_{i}^{des}$. Other word or sentence embedding techniques \cite{devlin2019bert, lu2020concept, lu2019graph} can also be used, such as BERT \cite{devlin2019bert} to derive items' content embeddings. Although these content embeddings capture rich semantic information, they fail to encode user--item interaction behaviors. To compensate for this limitation, we further incorporate collaborative embeddings into the tokenization process. Specifically, we derive item collaborative embeddings $\mathbf{e}_i^{col}$ from a pretrained SASRec model~\cite{kang2018self}, a representative sequential recommendation model that captures user behavior patterns and item dependencies. We then apply two separate MLPs to project the content and collaborative embeddings into a unified dimension:
\begin{equation}
\label{content_collaborative}
  \mathbf{z}_i^{con}=[\mathbf{e}_i^{tit}||\mathbf{e}_i^{des}]\mathbf{W}_1, \quad \mathbf{z}_i^{col}=\mathbf{e}_i^{col}\mathbf{W}_2,
\end{equation}
where $\mathbf{W}_1\in\mathbb{R}^{d_1\times d_2}$ and $\mathbf{W}_2\in\mathbb{R}^{d_3\times d_2}$ are learnable weight matrices. $||$ is the embedding concatenation operation. Dimension $d_1=2\times4096$ and $d_2=d_3=64$ in our implementation. Then, item $v_i$'s embedding is represented as $\mathbf{z}_i=\mathbf{z}_i^{con}||\mathbf{z}_i^{col}$. We then project item embedding into a latent space via an encoder for subsequent vector quantization:
\begin{equation}
\label{encoder}
  \mathbf{r}_i=\mathbf{Encoder}(\mathbf{z}_i),
\end{equation}
where $\mathbf{r}_i\in\mathbb{R}^{d}$. The encoder is instantiated with a three-layer MLPs following previous work \cite{wang2024learnable, shi2025llada}. Afterwards, we adopt RQ-VAE to hierarchically discretize the latent representation $\mathbf{r}_i$ into a sequence of tokens.
Specifically, RQ-VAE maintains $L$ codebooks $\{\mathcal{C}_{1}, \mathcal{C}_{2}, \dots, \mathcal{C}_{L}\}$, where each codebook $\mathcal{C}_{l} = \{\mathbf{e}_{l,k}\}_{k=1}^{K}$ contains $K$ learnable codeword embeddings. The quantization proceeds iteratively: at each level $l$, the closest codeword to the current residual is selected, and the residual is updated by subtracting the selected codeword:
\begin{equation}
\label{c_i_l}
    c_{i,l} = \arg\min_{k \in \{1,\dots,K\}}
    \left\| \mathbf{r}_{i,l} - \mathbf{e}_{l,k} \right\|_2^2,
\end{equation}
\begin{equation}
\label{r_l}
    \mathbf{r}_{i,l+1} = \mathbf{r}_{i,l} -
    \mathbf{e}_{c_{i,l}},
\end{equation}
where $\mathbf{r}_{i,1} = \mathbf{r}_i$ and
$c_{i,l}$ denotes the assigned token index at level $l$. After this process, the resulting token sequence $v_i$ is
$[c_{i,1}, c_{i,2}, \dots,
c_{i,L}]$. Finally, a decoder reconstructs the original embedding from the quantized representation:
\begin{equation}
\label{decoder}
    \hat{\mathbf{z}}_i = \textbf{Decoder}\left( \textstyle\sum_{l=1}^{L} \mathbf{e}_{c_{i,l}} \right).
\end{equation}

The decoder mirrors the encoder with a symmetrically reversed architecture. The standard RQ-VAE training objective consists of a reconstruction loss and a vector quantization loss:
\begin{equation}
\label{L_rq_vae}
  \mathcal{L}_{RQ-VAE}=\mathcal{L}_{Recon}+\mathcal{L}_{RQ},
\end{equation}
\begin{equation}
  \mathcal{L}_{Recon}=\left\| \mathbf{z}_i - \hat{\mathbf{z}_i} \right\|_2^2,
\end{equation}
\begin{equation}
\begin{aligned}
        \mathcal{L}_{RQ} =\textstyle\sum_{l=1}^{L}[&\left\|
    \text{sg}[\mathbf{r}_{i,l}] - 
    \mathbf{e}_{c_{i,l}} \right\|_2^2\\
    &+\beta\left\|
    \mathbf{r}_{i,l} - \text{sg}\left[
    \mathbf{e}_{c_{i,l}} \right] \right\|_2^2],
\end{aligned}
\end{equation}
where $\text{sg}[\cdot]$ denotes the stop-gradient operator and $\beta$ is a hyperparameter, which is set to 0.25 following the common practice \cite{rajput2023recommender, wang2024learnable, shi2025llada}.

However, current vector quantization techniques are agnostic to content credibility, leading to credibility entanglement discussed in \textbf{CH 1}. Specifically, the standard \textit{RQ-VAE optimizes purely for reconstruction fidelity, causing the codebook to organize entries along directions that minimize quantization error rather than along the credibility dimension.} As a result, credible and uncredible items whose embeddings differ in factuality but are close in reconstruction distance may still be assigned to the same codebook entry.

To address this, we introduce a credibility regularization that explicitly encourages the tokenizer to separate credible and uncredible items prior to code assignment. To mitigate objective conflict between the quantization loss and credibility regularization---as directly applying regularization on $\mathbf{r}_i$ would cause credible/uncredible items with distinct topical content to collapse toward each other in the embedding space (empirical evidence is provided in Appendix \ref{cred_head})---we first project $\mathbf{r}_i$ into a dedicated subspace via a two-layer MLP before computing the regularization: $\hat{\mathbf{r}}_i = \mathbf{MLP}(\mathbf{r}_i)$. Formally:
\begin{equation}
\label{L_cred}
  \begin{aligned}
    \mathcal{L}_{Cred} &= \frac{1}{|\mathcal{I}_{cre}|}\sum_{v_i\in\mathcal{I}_{cre}}\left\|\hat{\mathbf{r}}_{i}-\mathtt{Cen}(\hat{\mathbf{r}}_{i})\right\|_2^2\\
    &+\frac{1}{|\mathcal{I}_{unc}|}\sum_{v_j\in\mathcal{I}_{unc}}\left\|\hat{\mathbf{r}}_{j}-\mathtt{Cen}(\hat{\mathbf{r}}_{j})\right\|_2^2\\
    &-\left\|\mathtt{Cen}(\hat{\mathbf{r}}_{i})-\mathtt{Cen}(\hat{\mathbf{r}}_{j})\right\|_2^2,
  \end{aligned}
\end{equation}
where $\mathcal{I}_{cre}$ and $\mathcal{I}_{unc}$ are credible item set and uncredible item set. $\mathtt{Cen}(\hat{\mathbf{r}}_{i})=\frac{1}{|\mathcal{I}_{cre}|}\sum\limits_{v_i\in\mathcal{I}_{cre}}\hat{\mathbf{r}}_{i}$ and $\mathtt{Cen}(\hat{\mathbf{r}}_{j})=\frac{1}{|\mathcal{I}_{unc}|}\sum\limits_{v_j\in\mathcal{I}_{unc}}\hat{\mathbf{r}}_{j}$ are the centroids of the $\hat{\mathbf{r}}_{i}$ and $\hat{\mathbf{r}}_{j}$ for credible and uncredible items. The first two terms encourage items within each
credibility group to cluster tightly around their respective centroid (\emph{intra-class compactness}), while the third term pushes the two centroids apart (\emph{inter-class separation}), thus promoting a clear separation between credible and uncredible items in the tokenization space. The rationale for our centroid-based design is provided in Appendix \ref{complexity}.

The overall training objective of our designed credibility-aware tokenizer is:
\begin{equation}
\label{L_tokenization}
    \mathcal{L}_{Tokenization} = \mathcal{L}_{RQ-VAE}
    + \mathcal{L}_{Cred}.
\end{equation}

\subsection{Accuracy-preserving and Credibility-oriented Generator}
To address \textbf{CH2}, we propose an accuracy-preserving and credibility-oriented generator based on discrete diffusion. Specifically, we design an \textbf{asymmetric masking probability reduction strategy} to suppress the contributions of the last tokens of uncredible items during generation, while preserving valuable content-level topical signals.

Our design is motivated by the hierarchical nature of our credibility-aware tokenizer: the earlier levels capture dominant semantic patterns such as topics and writing styles, while the later levels progressively encode finer-grained details through residual refinement~\cite{rajput2023recommender}. Since \textit{factual discrepancies between credible and uncredible items typically manifest as subtle, detail-level differences} (e.g., specific claims, cited sources) rather than coarse topical divergence \cite{zhou2020survey, zhou2019fake, liu2026misspans, yang2023wsdms}, such credibility-discriminative signals are more likely to be preserved at the last few token levels after the shared topical information has been absorbed by the preceding levels. Conversely, reducing the masking probabilities of earlier levels would suppress tokens that encode valuable user preference signals, thereby degrading recommendation accuracy. Empirical evidence supporting this design choice is provided in Appendix \ref{layer_choice}.

Specifically, our generator operates through a forward masking process and a reverse denoising process. The forward process progressively masks tokens in the input sequence. For a masking ratio $r\sim U[0, 1]$, each token is independently masked with probability $r$. When $r=1$, all tokens are replaced with $\mathtt{[M]}$; when $r=0$, the sequence is fully observed. For the last token of uncredible item $v_i$, we further decrease the masking probability by a weight $w_i\in[0,1)$ to lower its contribution in the generation process. Formally, the masking probability of token $c_{i,l}^r$ at step $r$ is defined as:
\begin{equation}
    p_{\mathtt{M}}(c_{i,l}^r) = \begin{cases} 0,  & \text{if } c_{i,l}^r=\mathtt{[M]}\\
        w_i \cdot r, & \text{elif } v_i \in \mathcal{I}_{unc}\text{ and }l=L, \\
        r, & \text{else},
    \end{cases}
\end{equation}

This asymmetric masking probability reduction strategy suppresses the influence of uncredible signals during generation while leaving all other tokens---which encode valuable user preference signals such as topical interests---unaffected, thereby mitigating the inherent trade-off between recommendation accuracy and credibility.

This asymmetric reduction strategy is instantiated through two variants: \textbf{(1) Hard weighting strategy}, which sets $w_i = 0$ for the last tokens of all uncredible items, completely erasing their contributions to the generation process. \textbf{(2) Soft weighting strategy}, which assigns $w_i$ based on the credibility degree. Specifically, the credibility degree is estimated by computing the distance difference between the last token and the centroids of the credible and uncredible item clusters. Our motivation is: \textbf{the lower the credibility degree} of an item's last token---i.e., the closer it lies to the uncredible centroid---the more its associated information is contaminated by uncredible signals, and thus \textbf{the smaller the weight $w_i$ should be assigned to suppress its contribution}. Formally, $w_i$ in soft version is computed as:
\begin{equation}
\label{soft}
\sigma(\mathtt{D}(\mathbf{e}_{c_{i,L}},\, \mathtt{Cen}(\mathbf{e}_{c_{j,L}}))-\mathtt{D}(\mathbf{e}_{c_{i,L}},\, \mathtt{Cen}(\mathbf{e}_{c_{i',L}}))),
\end{equation}
where $\mathtt{D}(\cdot, \cdot)$ is the squared Euclidean distance and $\mathbf{e}_{c_{i,L}}$ is item $v_i$'s last token's embedding. $\mathtt{Cen}(\mathbf{e}_{c_{j,L}})$ and $\mathtt{Cen}(\mathbf{e}_{c_{i',L}})$ are centroids of last token embeddings for uncredible and credible items. $\sigma$ is the Sigmoid activation function.

At the reverse denoising step, a mask predictor $\mathcal{P}_{\theta}$ is trained to predict all masked tokens ($\mathtt{[M]}$) simultaneously. To mitigate the influence of uncredible items in both users' historical interactions and the target items, we design two training objectives: (1) \textbf{historical credibility loss} encourages the model to accurately predict the credible tokens in users' historical interactions, and (2) \textbf{target credibility loss} encourages the model to accurately predict the credible tokens in the target items. Formally, the training objectives are defined as:
\begin{equation}
\label{L_his}
    \mathcal{L}_{His} = -\mathbb{E}_{r}\left[\frac{1}{r}\sum_{c_{i,l}\in\mathcal{S}_{\mathcal{H}}}\log \mathcal{P}_{\theta}(c_{i,l}|\mathcal{S}_\mathcal{H}^r)\right],
\end{equation}
\begin{equation}
\label{L_tar}
    \mathcal{L}_{Tar} = -\mathbb{E}_{r}\left[\frac{1}{r}\sum_{l=1}^L\log \mathcal{P}_{\theta}(c_{n,l}|v_{n}^r, \mathcal{S}_\mathcal{H})\right],
\end{equation}
where \textbf{the sum is taken over masked positions}, i.e., $c_{i,l}^r=\mathtt{[M]}$ and $c_{n,l}^r=\mathtt{[M]}$. $\mathcal{P}_{\theta}$ is a bidirectional Transformer. The overall training objective of our designed generator is:
\begin{equation}
\label{L_generation}
  \mathcal{L}_{Generation}=\lambda \mathcal{L}_{His}+\mathcal{L}_{Tar},
\end{equation}
where $\lambda$ controls the strength of historical credibility training objective.

During inference, we begin with a fully masked token sequence $[\mathtt{M}, \cdots, \mathtt{M}]$ and iteratively apply the mask predictor $\mathcal{P}_\theta$ to decode the target item tokens over $T$ diffusion steps. At each step, the $\frac{L}{T}$ token positions with the highest predicted unmasking probabilities are decoded, while all remaining positions are re-masked to $\mathtt{[M]}$ and forwarded to the next step. During this process, we employ discrete diffusion-based beam search to obtain the top-$N$ most probable token sequences, which are subsequently mapped back to their corresponding items and returned as the final recommendations, following the practice of~\cite{shi2025llada}.

\section{Experiments}
\subsection{Datasets and Evaluation Metrics}
\noindent \textbf{Datasets.} Following prior work \cite{cai2026steering}, we conduct experiments on three real-world datasets: PolitiFact \cite{shu2020fakenewsnet}, \cite{shu2020fakenewsnet}, and MHMisinfo \cite{nguyen2025supporters}. The PolitiFact and GossipCop datasets are derived from FakeNewsNet repository\footnote{https://github.com/KaiDMML/FakeNewsNet} \citep{shu2020fakenewsnet}. PolitiFact and GossipCop datasets are user-news interaction datasets, where the uncredible items are fake news. MHMisinfo is a user-health information interaction dataset derived from a video-based mental health misinformation dataset\footnote{https://zenodo.org/records/13191247} \citep{nguyen2025supporters}, where the uncredible items are health misinformation. To the best of our knowledge,\textbf{ these are the only publicly available datasets that satisfy both requirements of this task}: user-item interaction data and item-level credibility labels. The statistics of these datasets are reported in Table \ref{tab:datasets}.

\begin{table}[t]
\small
\setlength{\tabcolsep}{3pt}
\caption{The statistics of the used three datasets.}
\begin{tabular}{lcccc}
\hline
Datasets & PolitiFact & GossipCop & MHMisinfo \\ \hline
\# Training sequences & 103,335 & 510,149 & 38,083 \\
\# Test sequences & 21,490 & 68,002 & 8,060 \\
\# Items & 616 & 9,529 & 3,160 \\
\# Credible items & 306 & 6,792 & 2,815 \\
\# Uncredible items & 310 & 2,737 & 345 \\ 
 \hline
\end{tabular}
\label{tab:datasets}
\end{table}

\begin{table*}[t]
\centering
\caption{Performance comparison of different methods across datasets. The \textcolor{red}{best} performance is in \textcolor{red}{red}, and the \textcolor{blue}{second-best} performance is in \textcolor{blue}{blue}. The improvement is statistically significant (paired t-test and p-value<0.05).}
\label{tab:results}
\resizebox{\textwidth}{!}{
\begin{tabular}{ll|c@{\hskip 3pt}c@{\hskip 3pt}c@{\hskip 3pt}c@{\hskip 3pt}|cccc|cc@{\hskip 3pt}c@{\hskip 7pt}c@{\hskip 6pt}c@{\hskip 3pt}}
\toprule
 & Metrics & GRU4Rec & SASRec & DreamRec & PreferDiff & Rec4Mit & HDInt & PRISM & \texttt{Disco} & TIGER & LETTER & LLaDA-Rec & \cellcolor{gray!20}\texttt{CreGR} (hard) & \cellcolor{gray!20}\texttt{CreGR} (soft) \\
\midrule
 \multirow{8}{*}{\rotatebox{90}{\textbf{PolitiFact}}} & HR@5 & 0.2142 & 0.2158 & 0.2416 & 0.2531 & 0.2039 & 0.2068 & 0.2043 & 0.2678 & 0.2109 & 0.2637 & \textcolor{blue}{0.2720} & \cellcolor{gray!20}0.2718 & \cellcolor{gray!20}\textcolor{red}{0.2762} \\
  & HR@10 & 0.3390 & 0.3519 & 0.3287 & 0.3554 & 0.3403 & 0.3413 & 0.3317 & 0.3775 & 0.3277 &  0.3838 & \textcolor{blue}{0.3916} & \cellcolor{gray!20}0.3875 & \cellcolor{gray!20}\textcolor{red}{0.3920} \\
  & NDCG@5 & 0.1463 & 0.1386 & 0.1767 & 0.1818 & 0.1348 & 0.1310 & 0.1299 & \textcolor{red}{0.1983} & 0.1439 & 0.1835  & 0.1942 & \cellcolor{gray!20}\textcolor{blue}{0.1957} & \cellcolor{gray!20}0.1949 \\
  & NDCG@10 & 0.1863 & 0.1823 & 0.2047 & 0.2147 & 0.1785 & 0.1741 & 0.1707 & \textcolor{red}{0.2336} & 0.1815 & 0.2252  & 0.2326 & \cellcolor{gray!20}\textcolor{blue}{0.2329} & \cellcolor{gray!20}0.2318 \\
  & CR@5 & 0.9266 & 0.9059 & 0.8620 & 0.8925 & 0.9118 & 0.9081 & 0.8888 & 0.9823 & 0.7994 &  0.9080 & 0.8887 & \cellcolor{gray!20}\textcolor{red}{0.9959} & \cellcolor{gray!20}\textcolor{blue}{0.9855} \\
  & CR@10 & 0.9122 & 0.9028 & 0.8437 & 0.8981 & 0.9061 & 0.9025 & 0.9150 & 0.9425 & 0.8046 & 0.8968  & 0.8691 & \cellcolor{gray!20}\textcolor{red}{0.9934} & \cellcolor{gray!20}\textcolor{blue}{0.9826} \\
  & HC@5 & 0.2930 & 0.2923 & 0.3096 & 0.3230 & 0.2818 & 0.2842 & 0.2799 & 0.3466 & 0.2761 & 0.3336 
  & 0.3374 & \cellcolor{gray!20}\textcolor{blue}{0.3517} & \cellcolor{gray!20}\textcolor{red}{0.3540} \\
  & HC@10 & 0.3889 & 0.3955 & 0.3695 & 0.3968 & 0.3887 & 0.3886 & 0.3846 & 0.4192 & 0.3612 & 0.4136 
 & 0.4120 & \cellcolor{gray!20}\textcolor{blue}{0.4354} & \cellcolor{gray!20}\textcolor{red}{0.4361} \\
\midrule
 \multirow{8}{*}{\rotatebox{90}{\textbf{GossipCop}}} & HR@5 & 0.2226 & 0.3078 & 0.4619 & 0.4969 & 0.2503 & 0.2154 & 0.2724 & 0.5236 & 0.5517 & 0.5106 & \textcolor{blue}{0.5803} & \cellcolor{gray!20}0.5605 & \cellcolor{gray!20}\textcolor{red}{0.5855} \\
  & HR@10 & 0.3194 & 0.4706 & 0.5501 & 0.6022 & 0.4054 & 0.3632 & 0.3920 & 0.6143 & 0.6254 & 0.6029 & \textcolor{blue}{0.6380} & \cellcolor{gray!20}0.6127 & \cellcolor{gray!20}\textcolor{red}{0.6384} \\
  & NDCG@5 & 0.1466 & 0.1607 & 0.3415 & 0.3655 & 0.1393 & 0.1141 & 0.1717 & 0.3996 & 0.4531 & 0.4017 & \textcolor{blue}{0.4930 } & \cellcolor{gray!20}0.4762 & \cellcolor{gray!20}\textcolor{red}{0.4984} \\
  & NDCG@10 & 0.1778 & 0.2135 & 0.3704 & 0.3999 & 0.1893 & 0.1618 & 0.2103 & 0.4292 & 0.4770 & 0.4317 & \textcolor{blue}{0.5118} & \cellcolor{gray!20}0.4932 & \cellcolor{gray!20}\textcolor{red}{0.5157} \\
  & CR@5 & 0.8864 & 0.8743 & 0.8464 & 0.8307 & 0.9087 & 0.9032 & 0.8384 & \textcolor{blue}{0.9277} & 0.8536 & 0.8418 & 0.8320 & \cellcolor{gray!20}\textcolor{red}{0.9601} & \cellcolor{gray!20}0.9132 \\
  & CR@10 & 0.8706 & 0.8526 & 0.8339 & 0.8228 & \textcolor{blue}{0.9065} & 0.8629 & 0.8790 & 0.9039 & 0.8439 & 0.8344 & 0.8236 & \cellcolor{gray!20}\textcolor{red}{0.9234} & \cellcolor{gray!20}0.8543 \\
  & HC@5 & 0.2964 & 0.3612 & 0.4417 & 0.4525 & 0.3228 & 0.2917 & 0.3032 & 0.4919 & 0.4813 & 0.4614 & 0.4846 & \cellcolor{gray!20}\textcolor{red}{0.5172} & \cellcolor{gray!20}\textcolor{blue}{0.5131} \\
  & HC@10 & 0.3685 & 0.4474 & 0.4744 & 0.4888 & 0.4280 & 0.3944 & 0.4144 & \textcolor{blue}{0.5208} & 0.5039 & 0.4931 & 0.5005 & \cellcolor{gray!20}\textcolor{red}{0.5266} & \cellcolor{gray!20}0.5118 \\
\midrule
 \multirow{8}{*}{\rotatebox{90}{\textbf{MHMisinfo}}} & HR@5 & 0.1151 & 0.1485 & 0.1819 & 0.1974 & 0.1480 & 0.1364 & 0.1424 & \textcolor{blue}{0.2215} & 0.1617  & 0.1780 & 0.1948 & \cellcolor{gray!20}0.2197 & \cellcolor{gray!20}\textcolor{red}{0.2240} \\
  & HR@10 & 0.1894 & 0.2592 & 0.2426 & 0.2620 & 0.2537 & 0.2164 & 0.2393 & \textcolor{blue}{0.2822} & 0.2110  & 0.2801 & 0.2422 & \cellcolor{gray!20}0.2742 & \cellcolor{gray!20}\textcolor{red}{0.2876} \\
  & NDCG@5 & 0.0760 & 0.0826 & 0.1313 & 0.1389 & 0.0874 & 0.0835 & 0.0969 & \textcolor{red}{0.1580} &  0.1131  & 0.1188 & 0.1396 & \cellcolor{gray!20}0.1397 & \cellcolor{gray!20}\textcolor{blue}{0.1480} \\
  & NDCG@10 & 0.0998 & 0.1179 & 0.1509 & 0.1598 & 0.1207 & 0.1094 & 0.1289 & \textcolor{red}{0.1778} & 0.1288  & 0.1513 & 0.1547 & \cellcolor{gray!20}0.1573 & \cellcolor{gray!20}\textcolor{blue}{0.1683} \\
  & CR@5 & 0.8380 & 0.8839 & 0.9002 & 0.8693 & 0.8641 & 0.8333 & 0.8667 & 0.9305  &  0.9118 & 0.9290 & 0.9117 & \cellcolor{gray!20}\textcolor{red}{0.9579} & \cellcolor{gray!20}\textcolor{blue}{0.9445} \\
  & CR@10 & 0.8608 & 0.8915 & 0.8952 & 0.8874 & 0.9056 & 0.8929 & 0.9212 & 0.9264 & 0.8757  & 0.9253 & 0.8914 & \cellcolor{gray!20}\textcolor{red}{0.9430} & \cellcolor{gray!20}\textcolor{blue}{0.9315} \\
  & HC@5 & 0.1806 & 0.2223 & 0.2591 & 0.2715 & 0.2205 & 0.2055 & 0.2144 & 0.3001 & 0.2387  & 0.2574 & 0.2730 & \cellcolor{gray!20}\textcolor{blue}{0.3012} & \cellcolor{gray!20}\textcolor{red}{0.3039} \\
  & HC@10 & 0.2630 & 0.3278 & 0.3147 & 0.3295 & 0.3252 & 0.2915 & 0.3150 & \textcolor{blue}{0.3507} & 0.2848  & 0.3489 & 0.3138 & \cellcolor{gray!20}0.3467 & \cellcolor{gray!20}\textcolor{red}{0.3556} \\
\bottomrule
\end{tabular}
}
\end{table*}

\noindent \textbf{Evaluation Metrics.} We adopt widely used metrics across three dimensions: recommendation accuracy (NDCG@$N$, HR@$N$), content credibility (CR@$N$, i.e., credible rate, the proportion of credible items in the recommendation list), and their balance (HC@$N$, which combines HR@$N$ and CR@$N$)~\cite{cai2026steering}. \textbf{All metrics are higher-the-better}, and their detailed formulations are provided in Appendix~\ref{app:metrics}.

\subsection{Baselines}
We compare our proposed model \texttt{CreGR} with the representative and state-of-the-art baselines from three categories: (i) \textbf{general recommendation models:} including two traditional models -- GRURec \cite{hidasi2015session}, SASRec \cite{kang2018self}, and two diffusion recommenders -- DreamRec \cite{yang2024generate} and PreferDiff \cite{liu2025preference}; (ii) \textbf{credible recommendation models:} Rec4Mit \cite{wang2022veracity}, HDInt \cite{wang2024hierarchical}, PRISM \cite{ma2025bridging},  \texttt{Disco} \cite{cai2026steering}; and (iii) \textbf{generative recommendation models:} TIGER \cite{rajput2023recommender}, LETTER \cite{wang2024learnable}, LLaDA-Rec \cite{shi2025llada}. The detailed introduction of these methods is provided in Appendix \ref{appendix:baselines}.
\subsection{Implementation Details}
For fair comparison, all baseline methods are initialized with hyperparameters suggested in their original papers and subsequently fine-tuned on our datasets to ensure their best 
performances are reported. We first train the tokenization stage with 10k epochs, and the learning rate is searched in range \{1e-5, 5e-5, 1e-4\}. We then train the generation stage and set the maximum epoch to 500, with learning rate searched in range \{5e-4, 1e-3, 3e-3, 5e-3\}. Batch size is searched in range \{256, 512, 1024, 2048, 4096\}. $L$, $K$, $T$ are set to 4, 256 and 4 following prior work \cite{shi2025llada}. $\lambda$ is searched in range \{3, 4, 5, 6, 7\}. Beam width is fixed to 50. Model parameters are optimized using AdamW \citep{loshchilov2017decoupled}. Each method is run five times, and the average performance is reported. More implementation details are provided in Appendix \ref{appendix:implementation}.

\subsection{Performance Comparison}
From the results reported in Table \ref{tab:results}, we have the following observations:

\noindent \textbf{Credibility Improvement.} \texttt{CreGR} achieves the best recommendation credibility (CR@5 and CR@10) across all datasets, demonstrating the effectiveness of our credibility-aware tokenization and accuracy-preserving and credibility-oriented generation designs. Notably, the hard weighting strategy (i.e., \texttt{CreGR} (hard)) attains higher credibility than its soft counterpart (i.e., \texttt{CreGR} (hard)), owing to its more aggressive suppression of uncredible token contributions.

\noindent \textbf{Accuracy Preservation.} \texttt{CreGR} preserves competitive recommendation accuracy. Although \texttt{CreGR} yields slightly lower NDCG than \texttt{Disco} on PolitiFact and MHMisinfo datasets, it maintains superior or comparable accuracy across all other cases. This can be attributed to: (1) our accuracy-preserving and credibility-oriented generator selectively reduces the masking probability of only the last token of uncredible items, leaving all other tokens that encode user preference signals unaffected; and (2) the incorporation of collaborative embeddings from a pretrained model further enriches item representations, providing additional accuracy gains.

\noindent \textbf{Credibility-accuracy Balance.} \texttt{CreGR} achieves the best performance in terms of HC@5 and HC@10, demonstrating its effectiveness in balancing recommendation accuracy and credibility. This achievement can be attributed to our designed accuracy-preserving and credibility-oriented generator which can mitigates uncredible content while preserving recommendation accuracy as much as possible. This balance is of particular practical significance, as both objectives are indispensable in real-world recommendation scenarios.

\begin{table*}[t]
\centering
\caption{Performance comparison between \texttt{CreGR} (full) and other model variants. The improvement is statistically significant (paired t-test and p-value<0.05).}
\label{tab:ablation}
\resizebox{\textwidth}{!}{
\begin{tabular}{l|cccc|cccc|cccc}
\toprule
Datasets & \multicolumn{4}{c|}{PolitiFact} & \multicolumn{4}{c|}{GossipCop} & \multicolumn{4}{c}{MHMisinfo} \\
Methods & HR@5 & NDCG@5 & CR@5 & HC@5 & HR@5 & NDCG@5 & CR@5 & HC@5 & HR@5 & NDCG@5 & CR@5 & HC@5 \\
\midrule
\cellcolor{gray!20}CreGR (full) & \cellcolor{gray!20}0.2762 & \cellcolor{gray!20}\textcolor{blue}{0.1949} & \cellcolor{gray!20}\textcolor{red}{0.9855} & \cellcolor{gray!20}\textcolor{red}{0.3540} & \cellcolor{gray!20}\textcolor{red}{0.5855} & \cellcolor{gray!20}\textcolor{red}{0.4984} & \cellcolor{gray!20}\textcolor{red}{0.9132} & \cellcolor{gray!20}\textcolor{red}{0.5131} & \cellcolor{gray!20}\textcolor{red}{0.2240} & \cellcolor{gray!20}\textcolor{blue}{0.1480} & \cellcolor{gray!20}\textcolor{red}{0.9445} & \cellcolor{gray!20}\textcolor{red}{0.3039} \\
w/o Title    & 0.2721 & 0.1930  & 0.9758 & 0.3494 & 0.5824 & 0.4952 & 0.8737 & 0.4992 & 0.2035 & 0.1461 & 0.9186 & 0.2820 \\
w/o Description     & 0.2732 & 0.1932 & 0.9380  & 0.3453 & 0.5838 & 0.4967 & \textcolor{blue}{0.8765} & \textcolor{blue}{0.5007} & 0.1993 & 0.1412 & 0.9154 & 0.2777 \\
w/o Collaborative     & \textcolor{red}{0.2819} & 0.1933 & 0.9446 & \textcolor{blue}{0.3531} & 0.5802 & 0.4953 & 0.8645 & 0.4954 & \textcolor{blue}{0.2192} & 0.1465 & \textcolor{blue}{0.9284} & \textcolor{blue}{0.2978} \\
w/o CT       & 0.2699 & 0.1914 & \textcolor{blue}{0.9768} & 0.3477 & 0.5785 & \textcolor{blue}{0.4971} & 0.8569 & 0.4923 & 0.2031 & 0.1413 & 0.9202 & 0.2818 \\
w/o APCOG    & \textcolor{blue}{0.2791} & \textcolor{red}{0.1979} & 0.8999 & 0.3445 & \textcolor{blue}{0.5850} & 0.4962 & 0.8424 & 0.4898 & 0.2153 & \textcolor{red}{0.1523} & 0.8302 & 0.2835 \\
\bottomrule
\end{tabular}
}
\end{table*}

\subsection{Ablation Study}
To evaluate the contribution of each component, we construct five variants and report results in Table~\ref{tab:ablation}. All results are based on \texttt{CreGR} (soft).

Variants \textbf{w/o Title}, \textbf{w/o Desc}, and \textbf{w/o Coll} remove title, description, and collaborative embeddings from the tokenization stage, respectively. The consistent performance drops confirm that each information source is indispensable: title and description provide semantic content signals, while collaborative embeddings supply user interaction patterns not captured by content alone.

Variant \textbf{w/o CT} replaces credibility-aware tokenizer with standard RQ-VAE tokenizer. The resulting degradation demonstrates that without explicit credibility disentanglement at the token level, the generation stage loses the ability to distinguish credible from uncredible content, jointly harming both recommendation accuracy and credibility.

Variant \textbf{w/o APCOG} disables accuracy-preserving and credibility-oriented generator by removing both hard and soft token weighting strategies. While this variant achieves marginal accuracy gains in certain cases, it incurs a substantial drop in recommendation credibility---an unacceptable trade-off given that credibility is a core optimization objective of \texttt{CreGR}.

\begin{figure}
    \centering
    \includegraphics[width=1\linewidth]{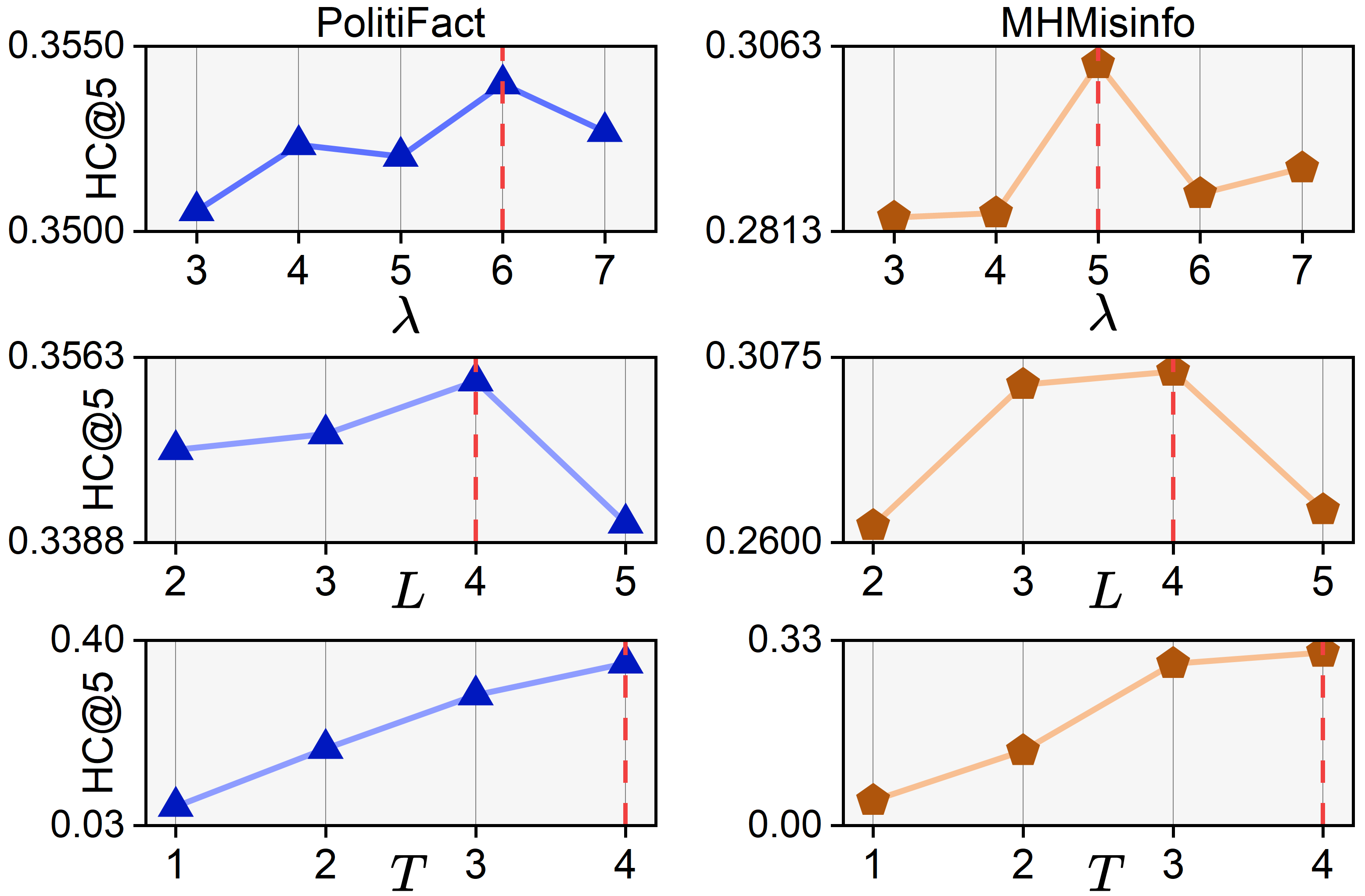}
    \caption{Effect of $\lambda$, $L$, $T$ on HC@5 of \texttt{CreGR} (soft).}
    \label{fig:lambda}
\end{figure}
\subsection{In-depth Analysis}
\subsubsection{Hyperparameters}
We vary $\lambda$ over $\{3, 4, 5, 6, 7\}$, $L$ over $\{2, 3, 4, 5\}$, and $T$ over $\{1, 2, 3, 4\}$ to examine their sensitivity. From Figure~\ref{fig:lambda}, we have following observations:

\noindent \textbf{(1) $\lambda$.} HC@5 first improves then degrades as $\lambda$ increases, peaking at $\lambda=6$ and $\lambda=5$ on PolitiFact and MHMisinfo. This suggests that a moderate $\lambda$ strikes the best balance between recommendation accuracy and credibility. Overly small or large values will degrades model performance.

\noindent \textbf{(2) $L$.} HC@5 first improves then degrades as $L$ increases, peaking at $L=4$ for all datasets. A larger $L$ provides finer-grained vector quantization but risks over-segmentation, leading to performance degradation beyond the optimal value. A smaller $L$ is sufficient to capture the hierarchical nature of item's semantic information, thereby harming model performance.

\noindent \textbf{(3) $T$.} HC@5 consistently rises as $T$ increases. Reducing the number of generation steps significantly increases the uncertainty of generation trajectory, and therefore decrease recommendation accuracy and credibility. Larger $T$ ensures finer-grained generation trajectories, leading to optimal recommendation performance.

Due to space constraints, results on GossipCop are omitted; this dataset exhibits a similar trend with optimal performance at $\lambda$=5, $L$=4 and $T$=4.

\subsubsection{Collision rate comparison}
\begin{table}[t]
\centering
\setlength{\tabcolsep}{4pt}
\small
\caption{Comparison of collision rate between generative recommendation methods.}
\label{tab:collision}
\begin{tabular}{lcccc}
\toprule
Datasets  & TIGER  & LETTER & LLaDA-Rec & \cellcolor{gray!20}\texttt{CreGR} \\
\midrule
PolitiFact & 8.0\% & 12.3\% & 8.0\% & \cellcolor{gray!20}0\% \\
GossipCop  & 5.9\%  & 5.3\%  & 5.0\% & \cellcolor{gray!20}0\% \\
MHMisinfo  & 10.5\% & 3.1\%  & 0.5\% & \cellcolor{gray!20}0.1\% \\
\bottomrule
\end{tabular}
\end{table}

\begin{table*}[t]
\centering
\caption{Performance comparison under different levels of noisy labels.}
\label{tab:label-noise}
\resizebox{\textwidth}{!}{%
\begin{tabular}{l cccc cccc cccc}
\toprule
\multirow{2}{*}{\textbf{Method}}
& \multicolumn{4}{c}{\textbf{PolitiFact}}
& \multicolumn{4}{c}{\textbf{GossipCop}}
& \multicolumn{4}{c}{\textbf{MHMisinfo}} \\
\cmidrule(lr){2-5}\cmidrule(lr){6-9}\cmidrule(lr){10-13}
& HR@5 & NDCG@5 & CR@5 & HC@5
& HR@5 & NDCG@5 & CR@5 & HC@5
& HR@5 & NDCG@5 & CR@5 & HC@5 \\
\midrule
TIGER     & 0.2109 & 0.1439 & 0.7994 & 0.2761 & 0.5517 & 0.4531 & 0.8536 & 0.4813 & 0.1617 & 0.1131 & 0.9118 & 0.2387 \\
LETTER    & 0.2637 & 0.1835 & 0.9080 & 0.3336 & 0.5106 & 0.4017 & 0.8418 & 0.4614 & 0.1780 & 0.1188 & 0.9290 & 0.2574 \\
LLaDA-Rec & 0.2720 & 0.1942 & 0.8887 & 0.3374 & 0.5803 & 0.4930 & 0.8320 & 0.4846 & 0.1948 & 0.1396 & 0.9117 & 0.2730 \\
\texttt{CreGR}     & 0.2762 & 0.1949 & 0.9855 & 0.3540 & 0.5855 & 0.4984 & 0.9132 & 0.5131 & 0.2240 & 0.1480 & 0.9445 & 0.3039 \\
\midrule
\multicolumn{13}{l}{\textit{Label-noise proportion}} \\
\midrule
5\% (\texttt{CreGR})  & 0.2733 & 0.1923 & 0.9803 & 0.3509 & 0.5846 & 0.4998 & 0.9100 & 0.5117 & 0.2209 & 0.1501 & 0.9380 & 0.3003 \\
10\% (\texttt{CreGR}) & 0.2720 & 0.1913 & 0.9801 & 0.3498 & 0.5836 & 0.4969 & 0.9095 & 0.5112 & 0.2168 & 0.1471 & 0.9347 & 0.2962 \\
15\% (\texttt{CreGR}) & 0.2717 & 0.1932 & 0.9780 & 0.3493 & 0.5819 & 0.4949 & 0.9068 & 0.5097 & 0.2124 & 0.1453 & 0.9335 & 0.2919 \\
\bottomrule
\end{tabular}%
}
\end{table*}

The collision rate measures the proportion of distinct items assigned identical token sequences, where a lower value indicates more discriminative tokenization. As shown in Table~\ref{tab:collision}, \texttt{CreGR} achieves the lowest collision rate 
across all datasets. We attribute this to two designs: (1) credibility-aware tokenizer encourages discriminative token representations for credible and uncredible items, reducing token sequence collisions between factually divergent items; and (2) collaborative embeddings enrich item representations with user interaction signals, enabling finer-grained distinctions between items with similar content.

\subsubsection{Label-noise Performance}
\label{label-noise}
In our Limitations section, we point out that exploring semi-supervised or unsupervised approaches is part of our future work, since in real-world scenarios, items' credibility labels may not be always available. These can serve as the solutions when credibility labels are noise or sparse. In the current version, we do not explicitly design a mechanism to resolve the label-noise issue. However, we conducted additional experiments to investigate the performance of CreGR in label-noise scenarios. Specifically, we randomly flipped the labels of a portion (5\%, 10\%, and 15\%) of uncredible items during training (at test time, correct labels are used to accurately evaluate model performance).

As shown in Table \ref{tab:label-noise}, \texttt{CreGR} also achieves superior results in the label-noise setting. Specifically, when the noise level is raised to 15\%, the recommendation credibility (CR@5) of \texttt{CreGR} only drops slightly, from 0.9855 to 0.9780, still remaining at a very high level (significantly outperforming LETTER's 0.9080). This indicates that although without specific designs, our method still have high potential to achieve great recommendation accuracy and credibility in scenarios where partial labels are noise or unavailable.

\subsection{Efficiency Analysis}
We compare the efficiency of \texttt{CreGR} (soft) against LLaDA-Rec, the only other discrete diffusion-based baseline. As shown in Table~\ref{tab:cost}, \texttt{CreGR} achieves higher tokenization efficiency, as LLaDA-Rec employs a multi-head VQ-VAE whose parallel multi-head operations incur substantially higher training overhead. In terms of generation efficiency, \texttt{CreGR} remains comparable to LLaDA-Rec, confirming that our designs introduce negligible computational cost during training and inference.

\begin{table}[t]
\centering
\scriptsize
\setlength{\tabcolsep}{1.5pt}
\caption{Tokenization and generation cost comparison.}
\label{tab:cost}
\begin{tabular}{l|cc|cc|cc}
\toprule
\multirow{2}{*}{Stages} & \multicolumn{2}{c|}{Tokenization} & \multicolumn{4}{c}{Generation} \\
                         & \multicolumn{2}{c|}{Total (minutes)} & \multicolumn{2}{c|}{Training/epoch (s)} & \multicolumn{2}{c}{Inference (s)} \\
\midrule
Methods    & LLaDA-Rec & \cellcolor{gray!20}\texttt{CreGR} & LLaDA-Rec & \cellcolor{gray!20}\texttt{CreGR} & LLaDA-Rec & \cellcolor{gray!20}\texttt{CreGR} \\
\midrule
PolitiFact & 55  & \cellcolor{gray!20}31 & 16.8 & \cellcolor{gray!20}17.5  & 117.1 &   \cellcolor{gray!20}117.6    \\
GossipCop  & 414 & \cellcolor{gray!20}75 & 189.0  & \cellcolor{gray!20}194.8 &   1006.0    &  \cellcolor{gray!20}1025.5   \\
MHMisinfo  & 160 & \cellcolor{gray!20}47 & 5.7  & \cellcolor{gray!20}5.9   & 43.0    &   \cellcolor{gray!20}43.2    \\
\bottomrule
\end{tabular}
\end{table}

\section{Conclusion}
In this paper, we propose a novel credible generative recommendation model, \texttt{CreGR}, to mitigate the recommendations of uncredible content in RSs. Specifically, we design a credibility-aware tokenizer to learn discriminative item tokens via credibility regularization. Afterwards, we propose a novel accuracy-preserving and credibility-oriented generator to reduce the contributions of tokens encoding uncredible information in the generation process. During this process, tokens representing users' preference signals remain unaffected to preserve high recommendation accuracy. Extensive experiments on three real-world datasets demonstrate the effectiveness of \texttt{CreGR} on improving recommendation credibility while preserving high recommendation accuracy.

\section*{Limitations}
While our proposed model demonstrates promising results in generating credible content recommendations, there are several limitations that warrant further investigation.

\textbf{Label availability.}  Our model relies on the availability of labeled data to identify credible and uncredible items during the tokenization and generation stage. In real-world scenarios, such labels may not always be available or may be noisy, which could affect the performance of our model. We plan to explore semi-supervised or unsupervised approaches to mitigate this limitation in future work. For example, Large Language Models (LLMs) \cite{lixin2025survey, zhang2026harnessing, gu2026survey} and LLM agents \cite{huang2024understanding, lu2026choosing, lu2026newborn, lu2026query} can serve as effective tools to handle semi-supervised and unsupervised scenarios, due to their powerful open-world knowledge.

\textbf{Content modality.} Our current model primarily focuses on textual content (e.g., news articles). Although MHMisinfo is a user-video interaction dataset, it doesn't provide the video metadata. It only contains the video titles and descriptions. However, in many recommendation scenarios, items may contain multimodal content (e.g., images, videos) that also contribute to their credibility. Future work could explore extending our model to incorporate multimodal information for a more comprehensive assessment of item credibility via multimodal LLMs \cite{li2026revealing, yangadapting, yang2026walking, yang2026turning, yang2026towards, zhou2025large} or other techniques.

\textbf{Dataset Scalability.} The datasets used in our experiments (GossipCop: 510,149 sequences; PolitiFact: 103,335 sequences) contain substantially more interaction sequences than commonly used Amazon benchmarks in GR \cite{mcauley2015image, hou2024bridging} (Sports: 35,598; Beauty: 22,363; Toys: 19,412), with GossipCop even containing approximately 2-3 times more sequences than larger Amazon datasets (Automotive: 193,651; Phones: 157,212; Tools: 240,799). This indicates that our used datasets provide a robust basis for validating the effectiveness and generalizability of \texttt{CreGR}. However, we acknowledge that the item vocabularies of these datasets are relatively limited, as they represent the only publicly available benchmarks that satisfy the dual requirements of user--item interaction data and item-level credibility labels; Amazon datasets, despite their scale, cannot be used as they provide no such credibility annotations. We leave the evaluation on larger-scale datasets as future work, and note that our method can be seamlessly deployed on them once such resources become available.

\section*{Ethical Considerations}
This paper aims to develop a credibility-aware generative recommendation model to serve societal good by mitigating the dissemination of uncredible content (e.g., fake news and misinformation) through recommender systems~(RSs). All datasets used in our experiments are derived from publicly available sources and contain no personally identifiable information.


\bibliography{custom}

\appendix

\section{Related Work}
\label{sec:appendix}
\subsection{Generative Recommendation}
Existing GR methods \cite{hou2025generative, rajput2023recommender, yang2026sparse} generally follow a two-stage design: (i) tokenization stage that represents each item with discrete tokens, and (ii) generation stage that generates a sequence of tokens to represent the recommended items \cite{rajput2023recommender, hou2025actionpiece, guo2026mlps, hou2026expressiveness, ding2026well, kim2026fusid, wei2025cofirec, zhong2025pctx}.

\noindent\textbf{Item Tokenization.} From the perspective of tokenization, existing methods can be broadly categorized into clustering-based and vector quantization-based approaches. Clustering-based methods organize items into hierarchical groups and represent each item with the cluster ID \cite{si2024generative, wang2024eager, lv2024semantic}. Vector quantization-based methods learn a shared codebook of discrete tokens and represent each item with the corresponding tokens in the codebook \cite{wang2025progressive, mei2025semantic, jia2026local, wang2026nezha, hou2023learning, wang2025generative, zheng2025enhancing}. For example, TIGER~\cite{rajput2023recommender} pioneered this direction by utilizing RQ-VAE to quantize item content embeddings into hierarchical semantic IDs. Subsequent work has sought to enrich the tokenization process: Some works ~\cite{wang2024learnable, liu2025best, wang2025empowering, ye2025dual} integrates collaborative signals from pre-trained recommendation models into vector quantization through contrastive alignment, and
LC-Rec~\cite{zheng2024adapting} aligns a large language model with semantic IDs via multi-task learning to improve the model's understanding of discrete tokens. SETRec \cite{lin2025order} learns order-agnostic identifiers to improve the performance of GR through LLMs.

Nevertheless, these methods follow a two-stage paradigm where the tokenizer and the downstream recommender are optimized separately, leading to objective misalignment between the tokenizer and generator. To bridge this gap, recent efforts have explored end-to-end joint optimization: ETEGRec~\cite{liu2025generative} proposes alternating optimization between the tokenizer and recommender with alignment losses, while DIGER~\cite{fu2026differentiable} introduces differentiable semantic IDs via Gumbel-Softmax to enable direct gradient flow from the recommender back to the tokenizer.

\noindent\textbf{Generation Paradigm.}
From the perspective of generation, existing GR models can be categorized into autoregressive (AR) models and non-autoregressive (NAR) models. Most existing methods adopt AR generation, which generate tokens sequentially from left to right \cite{rajput2023recommender, wang2024learnable, fang2026prism, qiao2026text}. At industrial scale, HSTU~\cite{zhai2024actions} introduces hierarchical sequential transducers with significant scalability, and OneRec~\cite{deng2025onerec} unifies retrieval and ranking within a single generative model with preference alignment. However, the AR paradigm suffers from two inherent limitations: (i) the sequential decoding process is time-consuming, as each token must wait for all preceding tokens to be generated, and (ii) errors made at earlier generation steps propagate irreversibly to subsequent tokens, resulting in error accumulation~\cite{shi2025llada}.

To address these limitations, NAR models have been proposed \cite{hou2025generating, shah2025masked, liu2026diffgrm, mu2026masked}. Inspired by discrete diffusion language models \cite{nie2026large, li2025survey, liu2026diffusion} and diffusion models in RSs \cite{lin2024survey, liu2025flow, cai2026sprint}, LLaDA-Rec~\cite{shi2025llada} introduces a parallel generation paradigm, where all tokens are generated simultaneously conditioned on bidirectional context. This design enables richer inter-token dependency modeling by attending to both preceding and following tokens.

Although GR has shown promising results, existing methods generally focus on improving recommendation accuracy while ignoring the credibility of the recommended content. To fill this gap, in this paper, we propose a novel GR model that explicitly incorporates content credibility into both the tokenization and generation stages.

\subsection{Credible Content Recommendation}
RSs play a crucial role in helping users quickly locate content they are interested in \cite{cai2022fg,cai2023adaptive,cai2023all,bing2024automated,yuan2025multibehavior, zhang2025mitigating, wang2020modelling, wang2022exploiting}. However, RSs have been widely recognized as significant contributors to the spreading of uncredible content, including fake news, rumor, and misinformation, across modern web platforms \cite{fernandez2020recommender, samuel2019online}. The majority of prior work in this area has centered on two directions: examining whether standard recommendation algorithms amplify uncredible content, and developing mitigation strategies.

\noindent\textbf{Empirical Analysis.}
The majority of prior work has centered on examining whether standard recommendation algorithms amplify uncredible content \cite{fernandez2021analysing, pathak2023understanding, hornig2024misinformation, pathak2024analyzing, pathak2024advancing, fernandez2020recommender}. For instance, Pathak et al.~\cite{pathak2023understanding} conduct a systematic evaluation showing that collaborative filtering algorithms disproportionately recommend misinformation to users who have previously engaged with such content, creating self-reinforcing
filter bubbles. While these studies provide valuable diagnostic insights, they remain at the analysis level and do not propose concrete algorithmic solutions.

\noindent\textbf{Mitigation Strategies.}
In contrast, a smaller body of work moves beyond analysis to develop novel mitigation strategies \cite{wang2022veracity, wang2024hierarchical, sallami2023trust, ma2025bridging, cai2026misinformation}. Wang et
al.~\cite{wang2022veracity, wang2024hierarchical} propose veracity-aware news recommendation frameworks that explicitly disentangle items' veracity embeddings from event embeddings and solely utilize event embeddings to reduce the exposure of fake news. Sallami et al.~\cite{sallami2023trust} introduce FANAR, a trust-based adaptation of collaborative filtering that excludes untrustworthy neighbors from the recommendation process. More recently, Cai et al.~\cite{cai2026steering} explores to steer diffusion models toward credible content recommendation through two designs: a disentangled diffusion model and the progressively enhancement of credible subspace projection.

Although these mitigation methods have demonstrated effectiveness, they are fundamentally built upon discriminative modeling paradigms that rely on computing embedding similarities between user
and item embeddings to make recommendations. As such, they are inherently incompatible with the emerging GR paradigm, which formulates the recommendation task as a sequence generation problem over discrete tokens rather than a matching problem over continuous embeddings. To bridge this gap, in this paper, we propose a novel credibility-aware GR model.

\section{More Experimental Details}
\subsection{Baseline Introduction}
\label{appendix:baselines}
 
We compare our proposed \texttt{CreGR} with baselines from three categories:
 
\noindent\textbf{General Recommendation Models}
\begin{itemize}[left=0pt]
    \item \textbf{GRU4Rec}~\citep{hidasi2015session}: A session-based recommendation model that employs Gated Recurrent Units (GRUs) to model sequential user interactions.
 
    \item \textbf{SASRec}~\citep{kang2018self}: A self-attentive sequential recommendation model that leverages the Transformer's self-attention mechanism to capture long-range dependencies in user interaction histories.
 
    \item \textbf{DreamRec}~\citep{yang2024generate}: A diffusion recommender that generates the target item embeddings by applying a continuous diffusion process in the embedding space.
 
    \item \textbf{PreferDiff}~\citep{liu2025preference}: A preference-guided diffusion model for sequential recommendation that incorporates user preference signals into the diffusion process.
\end{itemize}
 
\noindent\textbf{Credible Recommendation Models}

\begin{itemize}[left=0pt]
    \item \textbf{Rec4Mit}~\citep{wang2022veracity}: A veracity-aware and event-driven personalized news recommendation model designed to mitigate fake news in recommender systems.
 
    \item \textbf{HDInt}~\citep{wang2024hierarchical}: A hierarchical and disentangling interest learning framework for unbiased and true news recommendation.
 
    \item \textbf{PRISM}~\citep{ma2025bridging}: A diffusion model that bridges user interests and truthfulness for mitigating fake news exposure in recommendations.
 
    \item \textbf{ \texttt{Disco}}~\citep{cai2026steering}: A disentangled diffusion model that steers the diffusion process towards credible content recommendation.
\end{itemize}
 
\noindent\textbf{Generative Recommendation Models}
 
\begin{itemize}[left=0pt]
    \item \textbf{TIGER}~\citep{rajput2023recommender}: A generative retrieval model for recommender systems that represents each item with semantic IDs generated by RQ-VAE. It employs Transformer to generate the semantic ID tokens of the target item conditioned on the user's interaction history.
 
    \item \textbf{LETTER}~\citep{wang2024learnable}: A generative recommendation model that utilizes items' collaborative signals to guide the tokenization process.
 
    \item \textbf{LLaDA-Rec}~\citep{shi2025llada}: A non-autoregressive generative recommendation model based on discrete diffusion. It generates all tokens of the target item in parallel through a discrete diffusion process.
\end{itemize}
\subsection{Evaluation Metric Introduction}
\label{app:metrics}
 
We adopt the following evaluation metrics to comprehensively evaluate recommendation accuracy, content credibility, and their balance.
 
\subsubsection{Recommendation Accuracy Metrics}
 
\begin{itemize}[left=0pt]
    \item \textbf{HR@$N$ (Hit Rate at $N$)}: HR@$N$ measures whether the ground-truth item appears in the top-$N$ recommendation list. For a set of test users (i.e., equal to sequences) $\mathcal{U}$, it is defined as:
    \begin{equation}
        \text{HR@}N = \frac{1}{|\mathcal{U}|} \sum_{u \in \mathcal{U}} \mathbf{1}\left(v_u^{gt} \in \mathcal{R}_u^N\right),
    \end{equation}
    where $v_u^{gt}$ is the ground-truth next item for user $u$, $\mathcal{R}_u^N$ is the top-$N$ recommendation list for user $u$, and $\mathbf{1}(\cdot)$ is the indicator function.
 
    \item \textbf{NDCG@$N$ (Normalized  Discounted Cumulative Gain at $N$)}: NDCG measures the ranking quality of the recommendation list by assigning higher scores when the ground-truth item appears at higher positions. It is defined as:
    \begin{equation}
        \text{NDCG@}N = \frac{1}{|\mathcal{U}|} \sum_{u \in \mathcal{U}} \frac{1}{\log_2(\text{rank}_u + 1)},
    \end{equation}
    where $\text{rank}_u$ denotes the position of the ground-truth item $v_u^{gt}$ in the top-$N$ recommendation list $\mathcal{R}_u^N$. If the ground-truth item does not appear in the list, the NDCG contribution is 0.
\end{itemize}
 
\subsubsection{Content Credibility Metric}
 
\begin{itemize}[left=0pt]
    \item \textbf{CR@$N$ (Credible Rate at $N$)}~\citep{cai2026steering}: CR@$N$ measures the proportion of credible items in the top-$N$ recommendation list. A higher CR@$N$ indicates that the RSs exposes users to fewer uncredible items. It is defined as:
    \begin{equation}
        \text{CR@}N = \frac{1}{|\mathcal{U}|} \sum_{u \in \mathcal{U}} \frac{\left|\mathcal{R}_u^N \cap \mathcal{I}_{cre}\right|}{|\mathcal{R}_u^N|},
    \end{equation}
    where $\mathcal{I}_{cre}$ is the set of credible items and $\mathcal{R}_u^N$ is the top-$N$ recommendation list for user $u$. CR@$N$ ranges from 0 to 1, where 1 indicates all recommended items are credible.
\end{itemize}
 
\subsubsection{Combined Metric}
 
\begin{itemize}[left=0pt]
    \item \textbf{HC@$N$}~\citep{cai2026steering}: HC@$N$ is a combined metric that evaluates the balance between recommendation accuracy (HR@$N$) and content credibility (CR@$N$). It is defined as:
    \begin{equation}
        \text{HC@}N = \frac{2 \times \text{HR@}N \times (\frac{\text{CR@}N}{2})}{\text{HR@}N + (\frac{\text{CR@}N}{2})}.
    \end{equation}
    A high HC@$N$ score requires both high recommendation accuracy and high content credibility, making it a comprehensive indicator of overall recommendation quality. Specifically, CR@$N$ is scaled by a factor of $\frac{1}{2}$ to align its magnitude with HR@$N$, ensuring that neither metric disproportionately dominates the combined score.
\end{itemize}
 
We evaluate all metrics at $N \in \{5, 10\}$, following standard practices in sequential recommendation~\citep{kang2018self}.

\subsubsection{Collision Rate}
\textbf{Collision rate} measures the proportion of items that share identical SIDs (i.e., identical discrete token sequences) with other items:
\begin{equation}
    \text{Collision Rate} = 1 - \frac{|\mathcal{SID}_{\text{unique}}|}{|\mathcal{I}|},
\end{equation}
where $|\mathcal{SID}_{\text{unique}}|$ is the number of distinct SIDs and $|\mathcal{I}|$ is the total number of items. A lower collision rate indicates better codebook utilization and finer-grained item discrimination.

\subsection{More implementation  details}
\label{appendix:implementation}
\noindent\textbf{Item Content.} Items in PolitiFact and GossipCop datasets consist of news articles, for which we use titles and descriptions to derive content representations. In MHMisinfo dataset, although items are videos, only titles and descriptions are available; we therefore rely solely on these fields for content representation. We encode each item's title and description into dense embeddings using LLaMA2-7B~\cite{touvron2023llama}.

\noindent\textbf{Tokenization Stage.} The projection matrices $\mathbf{W}_1$ and $\mathbf{W}_2$ have dimensions $(8092, 64)$ and $(64, 64)$, respectively, each followed by layer normalization. The dimension of collaborative embeddings is 64. The resulting content and collaborative embeddings are concatenated to form the final item embedding. The $\mathbf{Encoder}(\cdot)$ follows the architecture $\{128\!\rightarrow\!512, \text{ReLU}, 512\!\rightarrow\!256, \text{ReLU}, 256\!\rightarrow\!64\}$, and the $\mathbf{Decoder}(\cdot)$  mirrors it symmetrically as $\{64\!\rightarrow\!256, \text{ReLU}, 256\!\rightarrow\!512, \text{ReLU}, 512\!\rightarrow\!128\}$. Prior to computing $\mathcal{L}_{Cred}$ (Equation~\ref{L_cred}), we pass $r_{i}$ through a two-layer MLP with architecture $\{64\!\rightarrow\!64, \text{ReLU}, 64\!\rightarrow\!64\}$, which mitigates objective conflict between the quantization loss and credibility regularization. Learning rate of the credibility-aware tokenizer is searched in range \{5e-5, 1e-4, 5e-4, 1e-3, 5e-3\}.

\noindent\textbf{Generation Stage.} We adopt a bidirectional Transformer as the mask predictor. The numbers of layers, attention heads, and KV heads are set to $\{2, 8, 8\}$, $\{6, 8, 8\}$, and $\{2, 8, 8\}$ for PolitiFact, GossipCop, and MHMisinfo, respectively. The beam width is fixed at 50 for all datasets.  For fair comparison, the embedding dimension $d$ is set to 64 for all methods except continuous diffusion-based baselines (DreamRec~\cite{yang2024generate}, PreferDiff~\cite{liu2025preference}, PRISM~\cite{ma2025bridging}, and \texttt{Disco}~\cite{cai2026steering}), which require an embedding dimension of 3072 to achieve competitive performance. This high embedding dimension requirement was also explicitly noted in prior work \cite{liu2025preference, cai2026steering}.

All experiments are conducted on an NVIDIA A40 GPU with 48GB of memory.

\section{Empirical results of removing interactions between users and uncredible items before training}
\label{simply_remove}
As mentioned in CH2, simply removing uncredible items from the training data and retraining the generation model inevitably degrades recommendation accuracy. Here we provide the empirical evidence. As shown in Table \ref{tab:cregr_ablation}, simply removing uncredible interactions before training significantly degrades recommendation accuracy (up to 23.53\%). This is because valuable information (e.g., topics) in the uncredible items that reflect users' preference is also removed.

\begin{table}[t]
\centering
\small
\setlength{\tabcolsep}{4pt}
\caption{Recommendation accuracy comparison between \texttt{CreGR} and simply removing uncredible interactions before training.}
\label{tab:cregr_ablation}
\resizebox{0.48\textwidth}{!}{
\begin{tabular}{lcccc}
\toprule
\textbf{Method} & \textbf{HR@5} & \textbf{HR@10} & \textbf{NDCG@5} & \textbf{NDCG@10} \\
\midrule
\multicolumn{5}{l}{PolitiFact} \\
\midrule
\texttt{CreGR}            & 0.2762  & 0.3920   & 0.1949  & 0.2318  \\
Simply removing  & 0.2319  & 0.3362  & 0.1661  & 0.1998  \\
\textit{Improv.} & -16.04\% & -14.23\% & -14.78\% & -13.81\% \\
\midrule
\multicolumn{5}{l}{GossipCop} \\
\midrule
\texttt{CreGR}            & 0.5855  & 0.6384  & 0.4984  & 0.5157  \\
Simply removing  & 0.4799  & 0.5201  & 0.4138  & 0.4269  \\
\textit{Improv.} & -18.04\% & -18.53\% & -16.97\% & -17.22\% \\
\midrule
\multicolumn{5}{l}{MHMisinfo} \\
\midrule
\texttt{CreGR}            & 0.2240  & 0.2876  & 0.1480   & 0.1683  \\
Simply removing  & 0.1713  & 0.2282  & 0.1152  & 0.1334  \\
\textit{Improv.} & -23.53\% & -20.65\% & -22.16\% & -20.74\% \\
\bottomrule
\end{tabular}
}
\end{table}

\section{Evidence supporting the design of our credibility-aware tokenizer}
\label{cred_head}
Prior to computing $\mathcal{L}_{\text{Cred}}$, we project $\mathbf{r}_i$ into a dedicated subspace via a two-layer MLP to obtain $\hat{\mathbf{r}}_i$, rather than applying credibility regularization directly on $\mathbf{r}_i$. This design choice is motivated by the need to mitigate objective conflict between the quantization loss and credibility regularization. Specifically, directly applying regularization on $\mathbf{r}_i$ would cause items with distinct topical content but the same credibility label to collapse toward each other in the embedding space, as this regularization directly pushes the latent representations of items with the same credibility level closer together regardless of their topical differences. This collapse would substantially increase the collision rate---the proportion of distinct 
items assigned identical token sequences---rendering the token representations insufficiently discriminative for accurate recommendation. To verify this, we conduct 
an ablation experiment comparing direct application of $\mathcal{L}_{\text{Cred}}$ on $\mathbf{r}_i$ against our MLP-projected $\hat{\mathbf{r}}_i$. As shown in Table~\ref{tab:cre_head}, bypassing the MLP projection leads to a significant rise in collision rate, empirically confirming the necessity of this design.

\begin{table}[H]
\centering
\setlength{\tabcolsep}{4pt}
\caption{Collision rate comparison between using $\mathbf{r}_i$ and $\hat{\mathbf{r}}_i$ for $\mathcal{L}_{\text{Cred}}$ on \texttt{CreGR} (soft).}
\begin{tabular}{lccc}
\toprule
 & PolitiFact & GossipCop & MHMisinfo \\
\midrule
$\mathbf{r}_i$ & 48.94\% & 93.87\% & 84.01\% \\
$\hat{\mathbf{r}}_i$ (ours) & 0\% & 0\% & 0.1\%\\
\bottomrule
\end{tabular}
\label{tab:cre_head}
\end{table}

\section{Evidence supporting the design of our accuracy-preserving and credibility-oriented generator}
\label{layer_choice}
Our accuracy-preserving and credibility-oriented generator selectively reduces the masking probabilities of the last tokens of uncredible items. This design is motivated by the hierarchical nature of RQ-VAE, whereby fine-grained factual information is more likely to be encoded in the last few tokens after coarser semantic patterns have been captured by the preceding levels. To validate this, we conduct experiments by varying the selected token position from layer 1 to layer 4 (ours). As shown in Table~\ref{tab:token_layer}, selecting tokens from earlier layers to reduce their masking probabilities leads to consistent degradation in recommendation credibility, confirming that fine-grained factual information is indeed concentrated in the last tokens. This validates the rationale of our design choice to apply masking probability reduction exclusively to the last tokens of uncredible items.

\begin{table}[H]
\centering
\caption{The effect of different selected token positions in the generation stage on recommendation credibility (CR@10) of \texttt{CreGR} (soft).}
\begin{tabular}{ccc}
\toprule
Token position & PolitiFact & MHMisinfo \\
\midrule
1 & 0.9241 & 0.9101\\
2 & 0.9598 & 0.9026\\
3 & 0.9741 & 0.9121\\
4 (ours) & 0.9826 & 0.9315\\
\bottomrule
\end{tabular}
\label{tab:token_layer}
\end{table}

\section{Case Study}
\begin{figure*}
    \centering
    \includegraphics[width=1\linewidth]{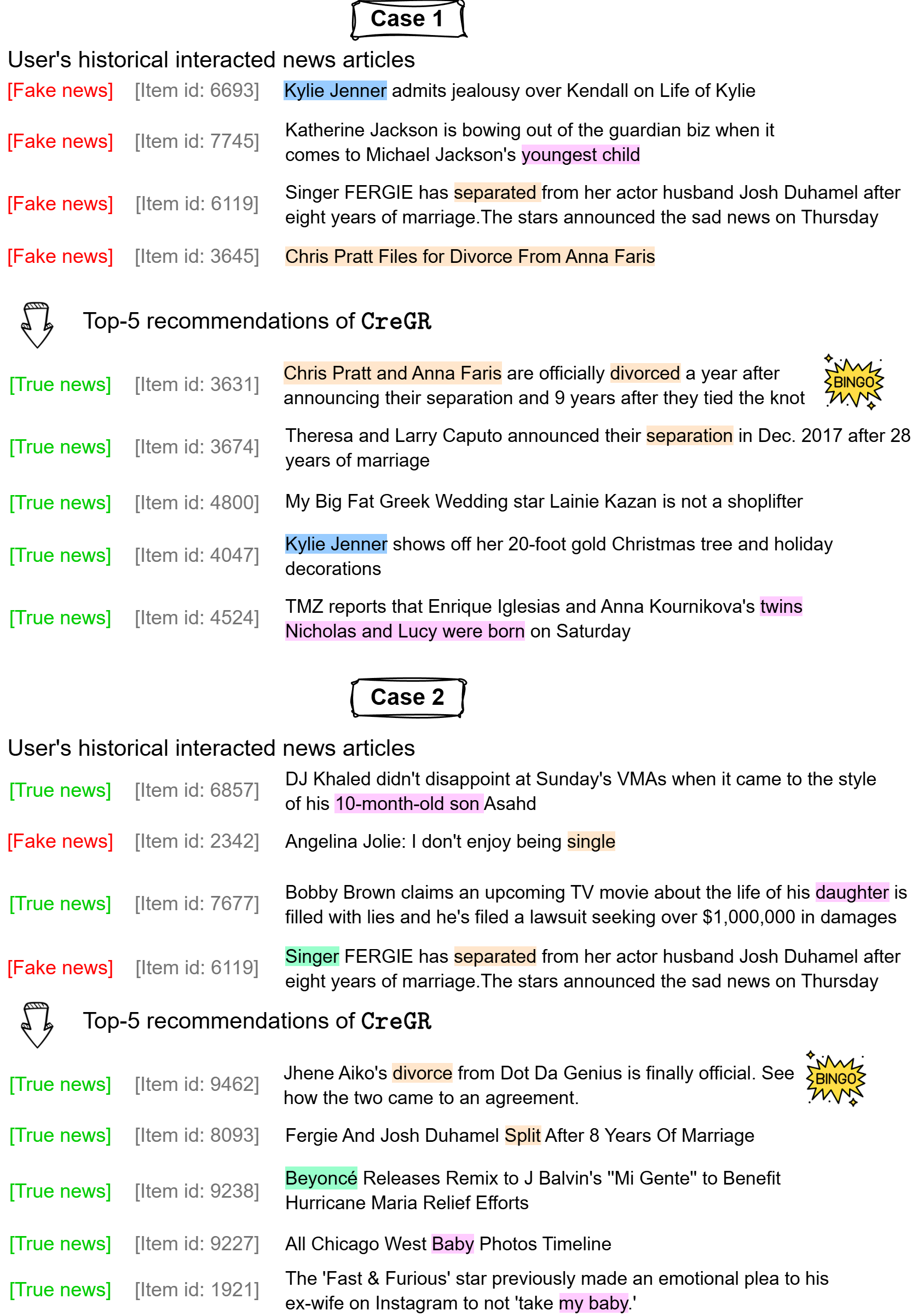}
    \caption{A case study illustrating the effectiveness of \texttt{CreGR} for mitigating uncredible items (fake news) on GossipCop dataset. Items sharing the same background color belong to the same topic cluster. “BINGO” indicates that the news articles is the one the user actually interacted with next.}
    \label{fig:case_study}
\end{figure*}
To intuitively illustrate the effectiveness of \texttt{CreGR}, we present a case study from the GossipCop dataset in Figure~\ref{fig:case_study}. In case 1, the user has historically
interacted with four fake news articles, covering topics such as celebrity relationships (e.g., ``Fergie separated from Josh Duhamel'', ``Chris Pratt files for divorce from Anna Faris'') and celebrity family affairs (e.g., ``Kylie Jenner admits jealousy over Kendall'', ``Katherine Jackson bowing out of guardian biz''). Despite all historical interactions being fake news, our model successfully captures the user's underlying topical preferences --- celebrity gossip, particularly regarding relationship changes and family dynamics --- while steering the recommendations entirely toward credible content. Specifically, the top-ranked recommendation (Item 3631) is a \emph{true news} article about ``Chris Pratt and Anna Faris are officially divorced'', which precisely matches the user's interest in the same celebrity couple but provides verified factual reporting instead of fabricated claims. Similarly, the second-ranked item (Item 3674) covers a celebrity separation topic that aligns with the user's preference pattern, and the fourth-ranked item (Item 4047) satisfies the user's interest in Kylie Jenner --- all being credible content. This case demonstrates that \texttt{CreGR} effectively mitigate uncredible content while still preserving users' preference-related information for improving recommendation accuracy.

\section{Complexity Analysis}
\label{complexity}
Compared to standard RQ-VAE quantization and discrete diffusion training, the additional computational overhead introduced by \texttt{CreGR} stems from two components: credibility regularization and the soft weighting strategy. For credibility regularization, the complexity of computing the distance between each item's latent representations and the centroids of credible and uncredible clusters is $O(|\mathcal{I}|\cdot d)$, where $|\mathcal{I}|$ is the number of items and $d$ is the codeword embedding dimension. For the soft weighting strategy, the complexity of identifying uncredible last tokens and computing $w_i$ consists of two terms: $O((|\mathcal{S}_{\mathcal{H}}|+|v_n|)\cdot L)$ for scanning token sequences, where $\mathcal{S}_{\mathcal{H}}$ and $v_n$ denote the token sequences of a user's historical interactions and the target item respectively, and $L$ is the codebook depth; and $O(|\mathcal{I}_{\text{unc}}|\cdot d)$ for computing weights over uncredible items, where $|\mathcal{I}_{\text{unc}}|$ is the number of uncredible items. All additional terms scale linearly with the input size and are negligible relative to the dominant cost of standard RQ-VAE and discrete diffusion training, confirming that \texttt{CreGR} maintains high computational efficiency.

\noindent \textbf{Computational efficiency of centroid-based credibility regularization.} A key design advantage of our credibility regularization (Eq. 11) is its computational efficiency. By computing distances to the centroids of credible and uncredible item clusters rather than to every individual item, we reduce the complexity of the credibility loss from $O(|\mathcal{I}|^2\cdot d)$ (pairwise) to $O(|\mathcal{I}|\cdot d)$ (centroid-based).

\begin{figure*}[t]
\centering
\begin{minipage}[t]{1\textwidth}
\begin{algorithm}[H]
\caption{Credibility-aware Tokenizer}
\label{alg:tokenization}
\begin{algorithmic}[1]
\Require Item set $\mathcal{V}$, credible item set $\mathcal{I}_{cre}$, uncredible item set $\mathcal{I}_{unc}$, initialized codebooks $\{\mathcal{C}_1, \ldots, \mathcal{C}_L\}$, the title, description, and collaborative embeddings of each item --- $\mathbf{e}_i^{tit}$, $\mathbf{e}_i^{des}$ and $\mathbf{e}_i^{col}$.
\Ensure Semantic ID (a sequence of discrete tokens) of each item: $[c_{i,1}, c_{i,2}, \cdots,c_{i,L}]$. Well-trained codebooks $\{\mathcal{C}_1, \ldots, \mathcal{C}_L\}$.
\For{each training epoch}
        \State Compute projected embeddings: $\mathbf{z}_i^{con} = [\mathbf{e}_i^{tit} \| \mathbf{e}_i^{des}]\mathbf{W}_1$, \ $\mathbf{z}_i^{col} = \mathbf{e}_i^{col}\mathbf{W}_2$ (Eq. \ref{content_collaborative})
        \State Form item embedding: $\mathbf{z}_i = \mathbf{z}_i^{con} \| \mathbf{z}_i^{col}$
        \State Encode: $\mathbf{r}_i = \mathbf{Encoder}(\mathbf{z}_i)$ (Eq. \ref{encoder})
        \State Initialize $\mathbf{r}_{i,1} = \mathbf{r}_i$
        \For{$l = 1$ to $L$}
            \State Assign token: $c_{i,l} = \arg\min_{k \in \{1,\ldots,K\}} \|\mathbf{r}_{i,l} - \mathbf{e}_{l,k}\|_2^2$ (Eq. \ref{c_i_l})
            \State Update residual: $\mathbf{r}_{i,l+1} = \mathbf{r}_{i,l} - \mathbf{e}_{c_{i,l}}$ (Eq. \ref{r_l})
        \EndFor
        \State Reconstruct: $\hat{\mathbf{z}}_i = \mathbf{Decoder}\left(\sum_{l=1}^L \mathbf{e}_{c_{i,l}}\right)$ (Eq. \ref{decoder})
    \State Compute RQ-VAE loss $\mathcal{L}_{RQ-VAE}$ (Eq. \ref{L_rq_vae}).
    \State Compute credibility regularization loss $\mathcal{L}_{Cred}$ (Eq. \ref{L_cred})
    \State Update parameters by minimizing $\mathcal{L}_{Tokenization} = \mathcal{L}_{RQ\text{-}VAE} + \mathcal{L}_{Cred}$ (Eq. \ref{L_tokenization})
\EndFor
\end{algorithmic}
\end{algorithm}
\end{minipage}
\hfill
\begin{minipage}[t]{1\textwidth}
\begin{algorithm}[H]
\caption{Accuracy-preserving and Credibility-oriented Generator}
\label{alg:generation}
\begin{algorithmic}[1]
\Require Token sequences $S_{\mathcal{H}}=[c_{1,1}, \cdots, c_{1,L}; \cdots; c_{n-1,1}, \cdots, c_{n-1,L}]$ of user interaction histories, target item tokens $v_n=[c_{n,1},\cdots,c_{n,L}]$, uncredible item set $\mathcal{I}_{unc}$, hyperparameter $\lambda$.
\Ensure Trained mask predictor $\mathcal{P}_\theta$.
\For{each training epoch}
        \State Sample masking ratio $r \sim U[0, 1]$
        \For{each token $c_{i,l}$ in $S_{\mathcal{H}}$ and $v_n$}
            \If {$c_{i,L}^r=\mathtt{[M]}$}
            \State Set masking probability $p_\mathtt{M}(c_{i,l}^r) = 0$
            \Else
            \If {$v_i \in \mathcal{I}_{unc}$ \textbf{and} $l = L$}
            \If {Hard weighting strategy}
            \State $w_i=0$
            \Else
            \State $w_i=\sigma(\mathtt{D}(\mathbf{e}_{c_{i,L}},\, \mathtt{Cen}(\mathbf{e}_{c_{j,L}}))-\mathtt{D}(\mathbf{e}_{c_{i,L}},\, \mathtt{Cen}(\mathbf{e}_{c_{i',L}})))$ (Eq. \ref{soft})
            \EndIf
                \State Set masking probability $p_\texttt{M}(c_{i,l}^r) = w_i \cdot r$ 
            \Else
                \State Set masking probability $p_\texttt{M}(c_{i,l}^r) = r$ 
            \EndIf
            \EndIf
            \State Mask $c_{i,l}$ with probability $p_\texttt{M}(c_{i,l}^r)$ to obtain $c_{i,l}^r$
        \EndFor
        \State Compute historical credibility loss $\mathcal{L}_{His}$ (Eq. \ref{L_his})
        \State Compute target credibility loss $\mathcal{L}_{Tar}$ (Eq. \ref{L_tar})
        \State Update $\mathcal{P}_\theta$ by minimizing $\mathcal{L}_{Generation} = \lambda \mathcal{L}_{His} + \mathcal{L}_{Tar}$ (Eq. \ref{L_generation})
\EndFor
\end{algorithmic}
\end{algorithm}
\end{minipage}
\end{figure*}

\section{Algorithm}
\label{app:algorithm}
We present the training procedure of \texttt{CreGR} in Algorithm~\ref{alg:tokenization} (tokenization stage) and Algorithm~\ref{alg:generation} (generation stage).

\renewcommand{\algorithmicrequire}{\textbf{Input:}}
\renewcommand{\algorithmicensure}{\textbf{Output:}}


\end{document}